\documentclass[aps,pra,12pt,showkeys,superscriptaddress,showpacs,reprint,raggedbottom,onecolumn]{revtex4-1}
\usepackage{amsmath}
\usepackage{amsthm}
\usepackage{graphicx}
\usepackage[colorlinks]{hyperref}
\allowdisplaybreaks
\usepackage{wrapfig}
\usepackage{cleveref}
\begin{document}
\title
{
Derivation of time-dependent transition probability for $2\mathrm{e}-2\mathrm{h}$ generation from 
$1\mathrm{e}-1\mathrm{h}$ state in the presence of external electromagnetic field
}
\date{\today} 
\author{Michael G. Bayne}
\author{Arindam Chakraborty}
\email[corresponding author: ]{archakra@syr.edu}
\affiliation{Department of Chemistry, Syracuse University, Syracuse, New York 13244 USA}
%--ABSTRACT--%
%\pacs{31.15.V}
\thispagestyle{plain}
\begin{center}
\begin{abstract}
	In this work, we investigate
the effect of electromagnetic (EM) field on the 
generation of 2e-2h states from 
1e-2h states. 
One of the fundamental ways
by which electromagnetic (EM) waves
interact with matter is by the generation of excited electronic states. 
The interaction of EM field 
with atoms and molecules is given by 
the field-dependent Hamiltonian. 
Excited states are
intrinsically transient in nature 
because they are not stationary
states of the field-dependent 
Hamiltonian.
Consequently, the time-dependent dynamics of excited 
states depend strongly on the external electromagnetic field.
Starting with the 1e-1h excitation
in a general many-electron system,
the system was propagated in time
using time-dependent perturbation theory (TDPT).
The expression for time-dependent
transition probability of 
$(1\mathrm{e}-1\mathrm{h}) \rightarrow (2\mathrm{e}-2\mathrm{h})$
was evaluated for a given time $t$
up to second-order in TDPT 
using diagrammatic techniques. 
The derivation 
does not assume any \textit{a priori} 
approximations to the
electron-electron correlation operator 
and presents the derivation of 
a complete set of contributing
diagrams associated with the 
full configuration interaction wave function.   
The result from this work show 
that the calculation of time-dependent 
transition probability
can be factored into a time-independent and time-dependent 
components. This is a significant outcome for 
efficient computation of the
time-dependent transition probability  
because it allows for 
pre-computation of time-independent 
components before the start of the calculations.
\end{abstract}
\end{center}
\maketitle

\section{Introduction}
%\input{sec_001_intro}
%\begin{itemize}
%\item Carrier multiplication (light harvesting solar cell beyond 100 percent efficiency) Art. Nozik.
%\cite{shockley_Detailed_1961}
%\cite{luo_Carrier_2008}
%\cite{mcguire_New_2008}
%\cite{padilha_Carrier_2013}
%\cite{stewart_Comparison_2012}
%\cite{padilha_Aspect_2013}
%\cite{beard_Multiple_2015}
%\item Multiexciton generation in QD  experiments (Klimov, Nozik, Beard...)
%\cite{shabaev_Multiexciton_2006}
%\cite{beard_Variations_2009}
%\cite{beard_Multiple_2008}
%\item Summary of theory work...(Prezhdo, Kilina, Jaeger, Tretiak, Li@UW....)
%\cite{beard_Third_2013}
%\cite{akimov_Advanced_2014}
%\cite{neukirch_Resolving_2014}
%\cite{jaeger_The_2012}
%\item Singlet fission in organic experimetal evidence
%\item Theory work (Van Vooris...)
%\cite{thompson_Magnetic_2015}
%\cite{wu_Singlet_2014}
%\cite{lee_Singlet_2013}
%\item Importance of electron-phonon coupling for MEG (theory)
%(Prezhdo and Jaeger, Vooris, Li...) 
%\cite{beard_Multiple_2011}
%\cite{hyeon_Photoexcited_2013}
%\cite{hyeon_Multiple_2012}
%\item Theory of EM-exciton interaction (Zunger, Xi) (PR papers)
%\cite{luo_Influence_2012}
%\end{itemize}
In 1961, Shockley and Queisser found the upper theoretical limit for the efficiency of p-n junction solar energy converters to be about 30\%. This is known as the Shockley-Queisser thermodynamic limit.\cite{shockley_Detailed_1961} Since then, there have been two main approaches for increasing the efficiency of the solar cell by means of producing multiple photogenerated excitons from a single absorbed photon. The two approaches are multiple exciton generation (MEG) (carrier multiplication (CM)) and singlet fission (SF).

In MEG, the exciton multiplication occurs when the absorbed photon is at least twice 
the nanocrystal band gap. This has been tested experimentally in semiconductor 
nanocrystals,\cite{shabaev_Multiexciton_2006,beard_Multiple_2008,beard_Variations_2009,
beard_Multiple_2015,stewart_Comparison_2012} quantum dots,\cite{nozik_Quantum_2002,beard_Third_2013,
schaller_High_2004,mcguire_New_2008} quantum wires, and quantum rods.\cite{padilha_Aspect_2013,beard_Quantum_2014} 
The affect of size, shape, and composition of PbS, PbSe, PbTe nanocrystals has on MEG was 
studied by Padilha et. al.\cite{padilha_Carrier_2013} MEG also has been shown to occur 
in carbon nanotubes\cite{gabor_Extremely_2009} as well as graphene.\cite{mcclain_Multiple_2010} 
The generation of multiexcitons has been subject of 
intense theoretical research.\cite{doi:10.1021/ar3002365}
For example, symmetry-adapted configuration 
interaction mehtod has been used to study the excited states of nanocrystals, such as lead selenide and silicon quantum dots, to determine the energetic threshold of MEG.\cite{jaeger_The_2012,akimov_Advanced_2014}
In addition to energetics requirements, 
the importance of electron-phonon coupling for multiexciton 
generation and multiexciton recombination (MER) in semiconductor quantum dots
has also been demonstrated.\cite{hyeon_Multiple_2012} 

The second avenue to generate multiple excitons is singlet fission. In molecular 
chromophores that have a triplet state energy that is close to 1/2 the energy of 
the first allowed optical transition (S$_{1}$-S$_{0}$), exciton multiplication can
occur upon photoexcitation to produce two triplet states from the single singlet 
state.\cite{johnson_The_2013,smith_Singlet_2010} Johnson et. al.  showed this using 
1,3-Diphenylisobenzofuran as a model chromophore.\cite{johnson_Singlet_2014} Thompson 
et. al. shows the magnetic field dependence of singlet fission in solutions 
of diphenyl tetracene.\cite{thompson_Magnetic_2015} Wu et. al. presents that tetracene 
is the best candidate in silicon solar cells to increase efficency using SF. They 
report a quantum efficiency of 127\% $\pm$ 18\%.\cite{wu_Singlet_2014} 
%\cite{lee_Singlet_2013} review article.

%\cite{beard_Multiple_2011}
%\cite{hyeon_Photoexcited_2013}

%Zunger et. al. investigates the vertical electric field tuning of the exciton fine-
%structure splitting (FSS) in several InGaAs and GaAs quantum dots using the atomistic 
%empirical pseudopotential approach and configuration interaction.\cite{luo_Influence_2012} 
%This work shows that the exciton FSS is highly tunable to the vertical electric field. 

In this work, we present a theoretical study of the effect of an external electromagnetic 
field on the generation of a biexcitonic state from a single excitonic state. 
The main goal of this work is to present a systematic 
derivation of the  
time-dependent transition probability for the $(1\mathrm{e}-1\mathrm{h}) \rightarrow (2\mathrm{e}-2\mathrm{h})$ process. 
We consider a general many-electron system in the presence of an 
external EM field.
The system is assumed to be excited at $t=0$ and the 
is propagated in time using field-dependent Hamiltonian. 
The form of the field-dependent Hamiltonian 
and the initial conditions  are described in \autoref{sec:sys_info}. 
The time-propagation of the state vector is performed
using time-ordered field-dependent propagator (\autoref{sec:time_prop}) 
using time-dependent perturbation theory
and the 0th, 1st and 2nd order contributions to the 
time-dependent transition amplitudes were 
derived in terms for second-quantized operators(\autoref{sec:tdpt}).
The transition amplitudes were expressed in terms 
of the time-independent Hugenholtz diagrams\cite{shavitt2009many} (\autoref{sec:diag})
with time-dependent vertex amplitudes. 
Finally, simplified expressions for calculating
time-dependent vertex amplitudes 
that is amenable to computer implementation
were derived (\autoref{sec:vertex}).
The key results and conclusions from the 
derivation are summarized in \autoref{sec:results}.
%\begin{itemize}
%    \item time-dependent MBPT....(Books: 
%    Fetter-Walecha\cite{fetter2003quantum}, 
%    Mahan\cite{mahan2000many}, 
%    March-Young\cite{march1967many}, 
%    Mattuck\cite{mattuck1976guide})
%    (Bartlett diagrams\cite{shavitt2009many})
%    (Atom-photon interaction Kohan tonogi book\cite{cohen1992quantum})
%\end{itemize}

\section{System information and definition}
\label{sec:sys_info}
We define the reference effective one-particle Hamiltonian as,
\begin{align}
 h_0 = \frac{-\hbar^2}{2m} \nabla^2 + v_\mathrm{ext} + v_\mathrm{eff}
\end{align}
where $v_\mathrm{eff}$ is the effective one-particle operator
and can be approximated using $v_\mathrm{HF}, v_\mathrm{KS}, v_\mathrm{ps}$,
or $v_\mathrm{model}$.
The eigenspectrum of the $h_0$ is used 
for the construction of the 
creation and annihilation operators
\begin{align}
	h_0 \chi_p = \epsilon_p \chi_{p}.
\end{align}
The N-electron non-interacting Hamiltonian
is defined as,
\begin{align}
	H_0 = \sum_{i}^N h(i).
\end{align}
The ground state of $H_0$ 
is defined as the quasiparticle vacuum,
\begin{align}
	\vert 0 \rangle \equiv \Phi_0.
\end{align}
The Hamiltonian for the 
interacting N-electron system 
is defined as, 
\begin{align}
	H = H_0 + W
\end{align}
where $W$ is the residual 
electron-electron interaction 
not included in the one-body 
operator $v_\mathrm{eff}$
\begin{align}
	W = \sum_{i<j}^N r_{ij}^{-1}  - \sum_{i}^N v_\mathrm{eff}(i).
\end{align}

The non-interacting electron-hole 
wave function is
defined using the creation operators for
quasi-electrons and quasi-holes
\begin{align}
	\vert \Phi_i^a \rangle
	&=
	\{a^\dagger i \} \vert 0 \rangle.
\end{align}
The correlated electron-hole wave function is defined 
using a correlation operator, $\Omega_n$,
\begin{align}
\label{eq:correlationOperator}
	\vert	\Psi\rangle
	&= \Omega_n \vert \Phi_i^a \rangle
\end{align}
where $\Omega_n$ will be defined later.

We are interested in the time-development of
the correlated wave function 
under the influence of an
external electromagnetic field.
The interaction between the molecule and the EM field
is given by the time-dependent interaction operator 
$V_F(t)$.\cite{cohen1992quantum} The total field dependent Hamiltonian
is defined as,
\begin{align}
	H_F(t)  = H_0 + V_F(t).
\end{align}

\section{Method for time-propagation}
\label{sec:time_prop}
In this work, we will work in the Dirac's 
interaction representation. 
In this representation, the total 
interaction potential is defined using the 
following similarity-transformation,
\begin{align}
	Z_I^F(t) 
	&=
	e^{+iH_0t/\hbar} [V_F(t)  + W] e^{-iH_0t/\hbar}. 
\end{align}
The field-dependent time-development operator, $U_F(t,0)$,
is defined as, 
\begin{align}
	U_F(t,0) = 1 + \sum_{n=1} U_F^{(n)}(t)
\end{align} 
where $U_F^{(n)}(t)$ is defined as,
\begin{align}
	U_F^{(n)}(t) 
	&= 
	C_n \int_0^t dt_1 dt_2 \dots dt_n 
	\mathcal{T}[Z_I^F(t_1) Z_I^F(t_2)\dots Z_I^F(t_n)].
\end{align}
We assume that the system at $t=0$ is described by 
the state vector $\Psi(0) = \Omega_n \vert \Phi_i^a \rangle$.
The time-development of this state vector to time $t$ 
is given by the following exprssion,
\begin{align}
	\vert \Psi_F(t) \rangle
	&=
	U_F(t,0) \vert \Psi(0) \rangle.
\end{align}
The subscript $F$ in the above equation implies that the 
time-development was performed under the influence of the 
the extenral field, $V_F$. 
In this work, we are interested in 
the 2e-2h generation from 1e-1h excitation.
\begin{align}
	\textrm{(Carrier multiplication) }
	P_{F,X \rightarrow X_2} (t) 
	&=
	\vert \langle 0 \vert \{ k^\dagger j^\dagger b c\}  \vert \Psi_F(t) \rangle \vert^{2}. 
\end{align}
For the purpose of this derivation, 
it is useful to write the transition probability
in terms of the transition amplitude $I$ as shown below,
\begin{align}
	P_{F,X \rightarrow X_2} (t_f) 
	&=
	\int_{0}^{t_f} dt \,
	[I_{F} (t)] [I_{F} (t)]^\ast 
\end{align}
where,
\begin{align}
	I_{F} (t)
	&=
\langle 0 \vert \{ k^\dagger j^\dagger b c\}  \vert \Psi_F(t) \rangle. 
\end{align}
In this work, we will use both  Wick's contraction 
and diagrammatic methods
for deriving the expression for the time-dependent transition amplitudes.
The first step in this many-step derivation is to write all the relevant quantities as vacuum expectation values. 
Writing the expression in terms of time-development operator,
\begin{align}
	I_F (t) &=
	\langle 0 \vert \{ k^\dagger j^\dagger b c\}  U_F(t,0)  \Omega_X \{ a^\dagger  i \}\vert 0 \rangle. 
\end{align}
For the nth-order term in the time-developmenet operator, we define
\begin{align}
	I_F^{(n)}(t_1,t_2,\dots,t_n)
	&=
	\langle 0 \vert \{ k^\dagger j^\dagger b c\}   \mathcal{T}[Z_I^F(t_1)Z_I^F(t_2)\dots Z_I^F(t_n)]  \Omega_X \{ a^\dagger  i \}\vert 0 \rangle.
\end{align}
Using Wick's theorem, we conclude the only fully contracted terms
will have non-zero contribution to the above expression
\begin{align}
\label{eq:I_F}
	I_F^{(n)}(t_1,t_2,\dots,t_n)
	&=
	\langle 0 \vert \{ k^\dagger j^\dagger b c\}  \mathcal{T}[Z_I^F(t_1)Z_I^F(t_2)\dots Z_I^F(t_n)]  \Omega_X \{ a^\dagger  i \}\vert 0 \rangle_{C}.
\end{align}
In this work, we evaluate the above expansion up to second-order
using diagrammatic techniques. The explicit expression for 
$I_F^{(0)}, I_F^{(1)}$ and $I_F^{(2)}$ are presented in 
Sec.~\ref{ssec:0orderContribution},~\ref{ssec:1orderContribution},and ~\ref{ssec:2orderContribution}. 

\section{Perturbative treatment of transition amplitudes}
\label{sec:tdpt}
\subsection{0th order contribution}
\label{ssec:0orderContribution}
The zeroth order term is field-independent and is given by the expression,
\begin{align}
		I_F^{(0)}
	&=
	\langle 0 \vert  \{k j^\dagger b c\} \Omega_X \{ a^\dagger  i \}\vert 0 \rangle_C.
\end{align}
As expected, the above expression is independent of time.
The Wick's contraction required to evaluate this 
term is denoted by the following expression,
\begin{align}
 \eta^{(3a)} 
 &=
 \langle 0 \vert  \{k j^\dagger b c\} \Omega_X \{ a^\dagger  i \}\vert 0 \rangle_L.
\end{align} 
We note that only connected diagrams contribute to the 
above expression and this fact is denoted by subscribe "L". 

\subsection{1st order contribution}
\label{ssec:1orderContribution}
The first-order term is:
\begin{align}
		I_F^{(1)}(t_1)
	&=
	\langle 0 \vert  \{k j^\dagger b c\} Z_I^F(t_1) \Omega_X \{ a^\dagger  i \}\vert 0 \rangle_C.
\end{align}
To evaluate the above expression, we will have to 
derive the expression of the the time-dependent interaction potential, $Z_I^F(t_1)$,
which is defined as,
\begin{align}
	Z_I^F(t) = e^{+iH_0t/\hbar} [V_F(t) + W]e^{-iH_0t/\hbar}.
\end{align}
In this derivation, we will split the above expression into
1-body and 2-body terms,
\begin{align}
	V_I^F(t) = e^{+iH_0t/\hbar}  V_F(t) e^{-iH_0t/\hbar}
\end{align}
\begin{align}
	W_I^F(t) = e^{+iH_0t/\hbar} W e^{-iH_0t/\hbar}.
\end{align}
The 1-body and 2-body time-dependent operators are
represented using time-dependent amplitudes,
\begin{align}
	V_I^F(t) 
	&=
	\sum_{pq} A_{pq}(t) p^\dagger q \\
	&=
	\sum_{pq} A_{pq}(t) \{p^\dagger q \} + \langle 0 \vert V_I^F(t) \vert 0 \rangle \\
	&=
	\sum_{pq} A_{pq}(t) \{p^\dagger q \} + \langle 0 \vert V_F(t) \vert 0 \rangle.
\end{align}
Similarly the 2-body term is given as,
\begin{align}
	W_I^F(t) 
	&=
	\frac{1}{2}
	\sum_{pqrs} B_{pqrs}(t) p^\dagger q^\dagger s r \\
	&=
	\frac{1}{2}\sum_{pqrs} B_{pqrs}(t) \{ p^\dagger q^\dagger s r \}
	+
	\sum_{pq} C_{pq}(t) \{ p^\dagger q \}
	+ \langle 0 \vert W_I^F(t)  \vert 0 \rangle 
\end{align}
where,
\begin{align}
	C_{pq}(t) = \sum_{i}^N B_{piqi}(t)-B_{piiq}(t).
\end{align}
Adding the terms and rewriting them in terms of 
normal-ordered 2-body, 1-body, and vacuum expectation value terms we get,
\begin{align}
	Z_I^F(t)  
	&=
	\frac{1}{2}\sum_{pqrs} B_{pqrs}(t) \{ p^\dagger q^\dagger s r \} 
    +	\sum_{pq} D_{pq}(t) \{p^\dagger q \} 
    + \langle 0 \vert Z_I^F(t) \vert 0 \rangle.
\end{align}
where,
\begin{align}
	\mathbf{D}(t)  &= \mathbf{A}(t) + \mathbf{C}(t) 
\end{align}
\begin{align}
	Z_I^F(t)
	&= 
	Z_{0}(t) + Z_{D}(t) + Z_{B}(t).
\end{align}
The 1st order probability for generation of 2e-2h from 1e-1h is given by the following expression,
\begin{align}
	I_{F}^{(1)} (t)
	&=
	 \langle 0 \vert  \{k j^\dagger b c\} [Z_0 + Z_D + Z_B] \Omega_X \{ a^\dagger  i \}\vert 0 \rangle_{C}.  
\end{align}
Summing over
\begin{align}
	I_{F}^{(1)} (t)
	=
	Z_0 (t) I^{(0)}  
	+ \sum_{pq} D_{pq}(t) \eta^{(4a)}_{pq}  
	+ \sum_{pqrs} B_{pqrs}(t) \eta^{(4b)}_{pqrs}
\end{align}
where,
\begin{align}
	\eta^{(4a)}_{pq}
	&=
	\langle 0 \vert  \{k j^\dagger b c\} 
	\{ p^\dagger q \}	
	 \Omega_X \{ a^\dagger  i \}\vert 0 \rangle_C  \\
	\eta^{(4b)}_{pqrs}
	&=
	\langle 0 \vert  \{k j^\dagger b c\}
	 \{ p^\dagger q^\dagger s r \}	
	 \Omega_X \{ a^\dagger  i \}\vert 0 \rangle_{C}. 
\end{align}

\subsection{2nd order contribution}
\label{ssec:2orderContribution}
The second-order term for ($t_1 > t_2$) is:
\begin{align}
 I_{F}^{(2)} (t)
	&=
	\langle 0 \vert  \{k j^\dagger b c\} Z_I^F(t_1) Z_I^F(t_2)  \Omega_X \{ a^\dagger  i \}\vert 0 \rangle_C.
\end{align}
Substituting,
\begin{align}
	Z_I^F(t_1) Z_I^F(t_2)
	&=
	[Z_0(t_1)+Z_D(t_1)+Z_B(t_1)][[Z_0(t_2)+Z_D(t_2)+Z_B(t_2)] \\
	&=
	     Z_0(t_1) [[Z_0(t_2)+Z_D(t_2)+Z_B(t_2)]  \notag \\ 
	&+ Z_D(t_1) [Z_0(t_2)+Z_D(t_2)+Z_B(t_2)] \notag \\ 
	&+ Z_B(t_1) [Z_0(t_2)+Z_D(t_2)+Z_B(t_2)]  \\
	&=
	     Z_0(t_1) [Z_0(t_2)+Z_D(t_2)+Z_B(t_2)]  \notag \\ 
	&+ [Z_D(t_1) + Z_B(t_1)] Z_0(t_2)  \notag \\
	&+ [Z_D(t_1)Z_B(t_2)]  + [Z_B(t_1) Z_D(t_2)]  \notag \\
	&+ [Z_D(t_1)Z_D(t_2)] + [Z_B(t_1) Z_B(t_2)].
\end{align}
Adding and subtracting $Z_0(t_1)Z_0(t_2)$
in the following expression,
\begin{align}
	[Z_D(t_1) + Z_B(t_1)] Z_0(t_2)
	&=
	[Z_0(t_1) + Z_D(t_1) + Z_B(t_1)] Z_0(t_2) - Z_0(t_1)Z_0(t_2).
\end{align}
Therefore,
\begin{align}
	Z_I^F(t_1) Z_I^F(t_2)
	&=
	     Z_0(t_1) [Z_0(t_2)+Z_D(t_2)+Z_B(t_2)]  \notag \\ 
	&+ [Z_0(t_1) + Z_D(t_1) + Z_B(t_1)] Z_0(t_2)  \notag \\
	&+ [Z_D(t_1)Z_B(t_2)]  + [Z_B(t_1) Z_D(t_2)]  \notag \\
	&+ [Z_D(t_1)Z_D(t_2)] + [Z_B(t_1) Z_B(t_2)] - [Z_0(t_1)Z_0(t_2)].
\end{align}
We define time-reversed anti-commutation as,
\begin{align}
	[A(t_1),B(t_2)]_{+}^{t} = A(t_1)B(t_2) + B(t_1)A(t_2).
\end{align}

Using the above equation, the expression for $I_{F}^{(2)} (t)$
is given as,
\begin{align}
\label{eq:amplitudes}
 I_{F}^{(2)} (t_1 t_2)
	&= 
	Z_0(t_1) I_{F}^{(1)} (t_2) + I_{F}^{(1)} (t_1)Z_0(t_2)
	- Z_0(t_1)Z_0(t_2) I^{(0)}  \notag\\
	&+ \langle 0 \vert  \{k j^\dagger b c\} [Z_D(t_1),Z_B(t_2)]^t_{+}  \Omega_X \{ a^\dagger  i \}\vert 0\rangle \notag \\
	&+ \langle 0 \vert  \{k j^\dagger b c\} [Z_D(t_1)Z_D(t_2)]  \Omega_X \{ a^\dagger  i \}\vert 0\rangle \notag \\
	&+ \langle 0 \vert  \{k j^\dagger b c\} [Z_B(t_1) Z_B(t_2)]  \Omega_X \{ a^\dagger  i \}\vert 0\rangle.
\end{align}
The evaluation of the terms in Eq.~\ref{eq:amplitudes} are given by,
\begin{align}
\langle 0 \vert  \{k j^\dagger b c\} [Z_D(t_1) Z_D(t_2)]  \Omega_X \{ a^\dagger  i \}\vert 0\rangle = 
\sum_{pq} \sum_{rs}  
 D_{pq}(t_1)D_{rs}(t_2) \eta_{pqrs}^{(5a)}
\end{align}
\begin{align}
\langle 0 \vert  \{k j^\dagger b c\} [Z_D(t_1),Z_B(t_2)]^t_{+}  \Omega_X \{ a^\dagger  i \}\vert 0\rangle = 
\sum_{pqrs} \sum_{xy} G_{pqrsxy}(t_1,t_2) \eta_{pqrsxy}^{(5b)}  
\end{align}
\begin{align}
\langle 0 \vert  \{k j^\dagger b c\} [Z_B(t_1) Z_B(t_2)]  \Omega_X \{ a^\dagger  i \}\vert 0\rangle = 
\sum_{pqrs} \sum_{tuvw}  
B_{pqrs}(t_1)B_{tuvw}(t_2) \eta_{pqrstuvw}^{(5c)}
\end{align}
where the time-independent components are given as,
\begin{align}
\eta_{pqrs}^{(5a)}
&=
\langle 0 \vert  \{k j^\dagger b c\} \{p^\dagger q\} \{r^\dagger s\} \Omega_X \{ a^\dagger  i \}\vert 0\rangle_C \\
\eta_{pqrsxy}^{(5b)}
&=
\langle 0 \vert  \{k j^\dagger b c\} \{p^\dagger q^\dagger s r\} \{x^\dagger y\} \Omega_X \{ a^\dagger  i \}\vert 0\rangle_C \\
\eta_{pqrstuvw}^{(5c)}
&=
\langle 0 \vert  \{k j^\dagger b c\} \{p^\dagger q^\dagger s r\} \{t^\dagger u^\dagger w v \} \Omega_X \{ a^\dagger  i \}\vert 0\rangle_C
\end{align}
and 
\begin{align}
G_{pqrsxy}(t_1,t_2) 
&=
B_{pqrs}(t_1)D_{xy}(t_2) + D_{xy}(t_1)B_{pqrs}(t_2).
\end{align}
\begin{align}
 I_{F}^{(2)} (t_1 t_2)
	&= 
	Z_0(t_1) I_{F}^{(1)} (t_2) + I_{F}^{(1)} (t_1)Z_0(t_2)
	- Z_0(t_1)Z_0(t_2) I^{(0)}  \notag\\
&+ \sum_{pqrs}      D_{pq}(t_1)D_{rs}(t_2) \eta_{pqrs}^{(5a)}
+ \sum_{pqrsxy}  G_{pqrsxy}(t_1,t_2) \eta_{pqrsxy}^{(5b)}  
+ \sum_{pqrstuvw} B_{pqrs}(t_1)B_{tuvw}(t_2) \eta_{pqrstuvw}^{(5c)}.
\end{align}

\section{Diagrammatic evaluation of Wick's contraction}
\label{sec:diag}
\begin{figure*}
  \begin{center}
      % left bottom right top
      %\fbox{\includegraphics[trim=4cm 20cm 4cm 2.5cm,scale=1.0]{./diagrams_ex_dynamics/table_diag3/table3.pdf}}  \\
      \includegraphics[trim=4cm 20cm 4cm 2.5cm,scale=1.0]{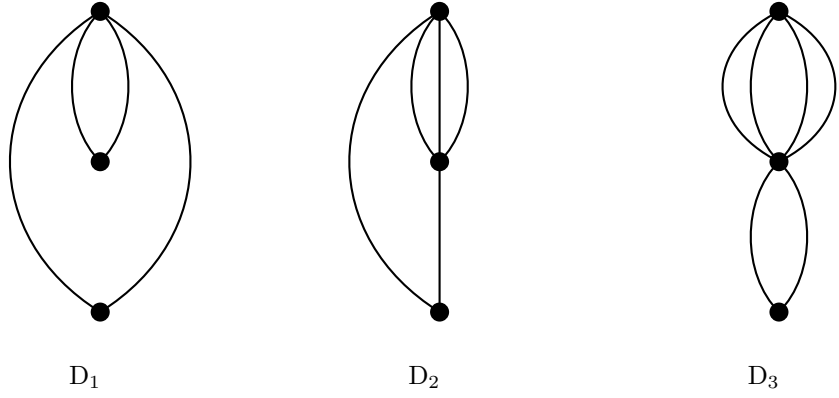}  \\
      \caption{\textrm{3-vertex diagams.}}
    \label{fig:3vert}
  \end{center}
\end{figure*}

\begin{figure*}
  \begin{center}
      % left bottom right top
      %\fbox{\includegraphics[trim=2.5cm 9cm 1.5cm 2.5cm,scale=1.0]{./diagrams_ex_dynamics/table_diag3/table4_3.pdf}}  \\
      \includegraphics[trim=2.5cm 9cm 1.5cm 2.5cm,scale=1.0]{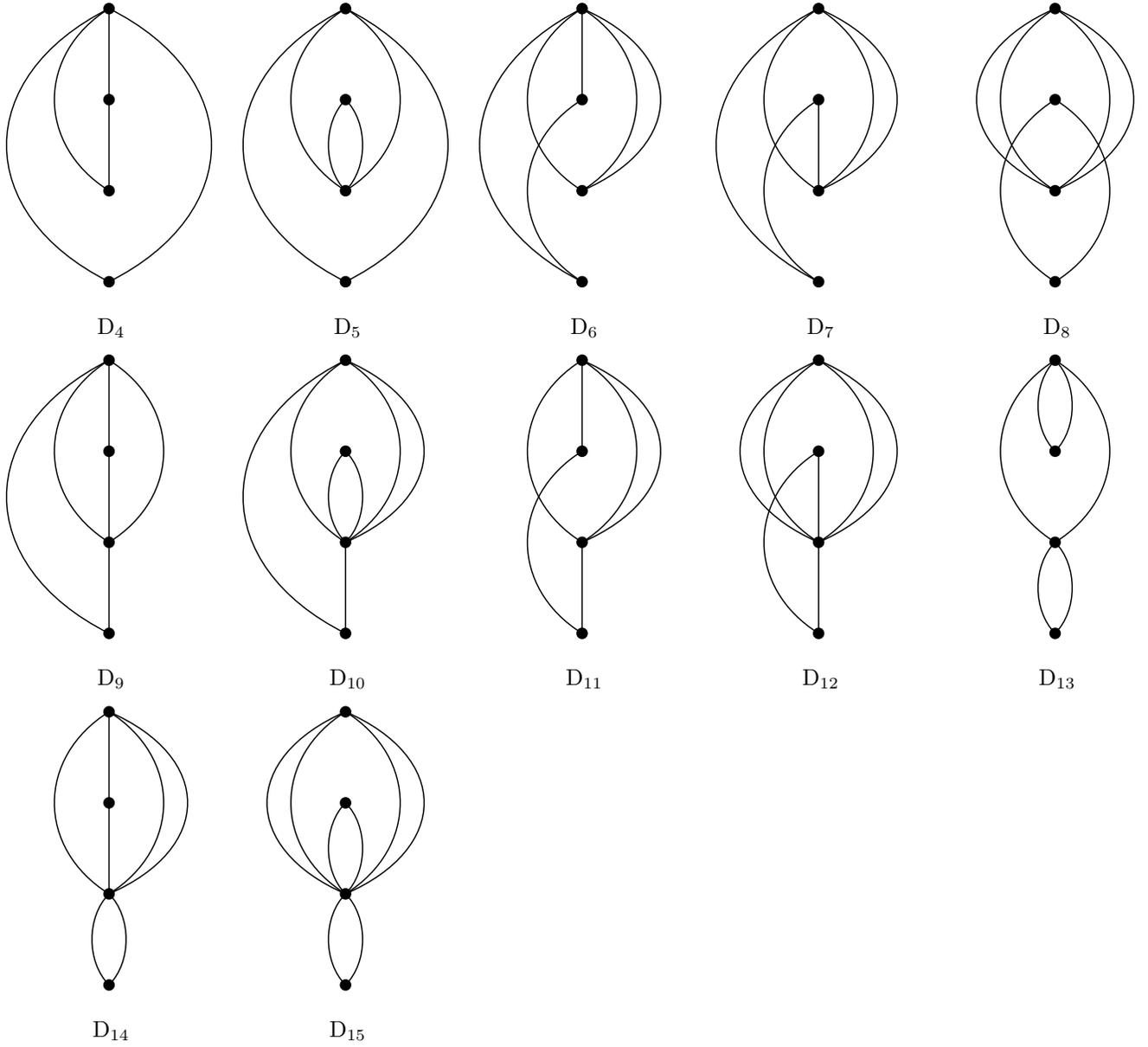}  \\
      \caption{\textrm{Part A: 4-vertex diagams.}}
    \label{fig:4verta}
  \end{center}
\end{figure*}

\begin{figure*}
  \begin{center}
      % left bottom right top
      %\fbox{\includegraphics[trim=2.5cm 3cm 1.5cm 2.5cm,scale=1.0]{./diagrams_ex_dynamics/table_diag3/table4.pdf}}  \\
      \includegraphics[trim=2.5cm 3cm 1.5cm 2.5cm,scale=1.0]{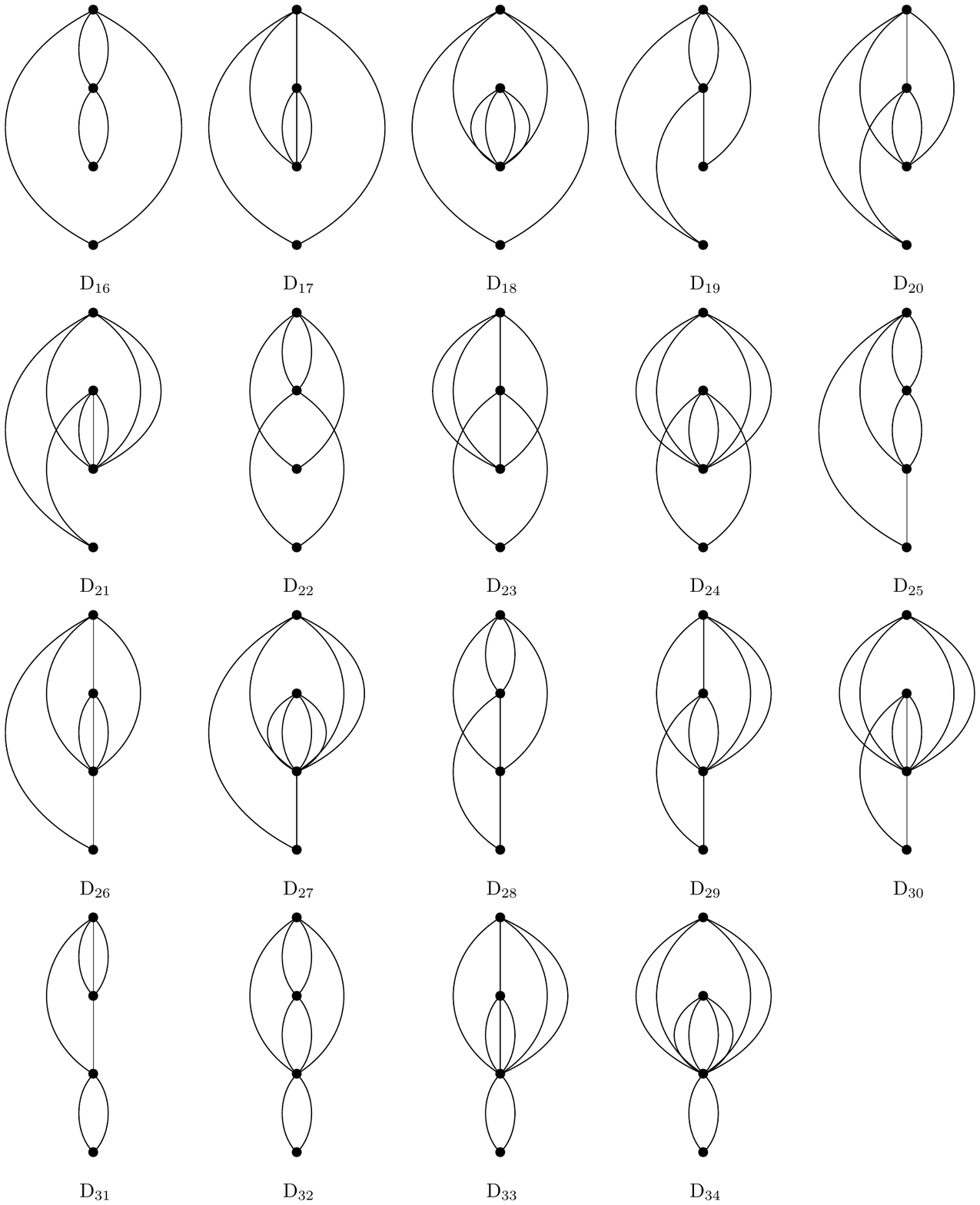}  \\
      \caption{\textrm{Part B: 4-vertex diagams.}}
    \label{fig:4vertb}
  \end{center}
\end{figure*}

In this section, we derive the expressions for the 
$\eta$ terms that are needed to 
evaluate the expression. 
The 3-vertex terms $\eta^{(3a)}$ are given by the 
set of diagrams presented in \autoref{fig:3vert}. We note that only linked-diagrams 
have non-zero contribution to $\eta^{(3a)}$.
The expression for $\eta^{(4)}$ can be
expressed as a sum of both linked and unlinked diagrams. 
However, it can be shown that all unlinked diagrams
have zero contribution. Analysis of the 
unlinked diagrams reveal that the 
unlinked diagrams contain the following expressions,
\begin{align}
	\langle 0 \vert \{ k^\dagger j^\dagger b c\} \{ a^\dagger i \} \vert 0 \rangle 
	\langle 0 \vert Z_{D,B} \Omega \vert 0 \rangle
	&= 0.
\end{align}
The set of linked diagrams for $\eta^{(4a)}$ 
and $\eta^{(4b)}$ are presented in Fig.~\ref{fig:4verta} and Fig.~\ref{fig:4vertb}.

The evaluation of the $\eta^{(5)}$ 
expressions require both linked and unlinked diagrams. 
In many cases, the unlinked 5-vertex diagrams
can be expressed in terms of the 3-vertex and 4-vertex 
diagrams derived earlier. 
In case of $\eta^{(5a)}$, this diagrammatic factorization 
is expressed as,
\begin{align}
	\eta^{(5a)}_{pqrs}
	&=
	\eta^{(2a)}_{pqrs} \eta^{(3a)} + \eta^{(5aL)}_{pqrs}
\end{align}
where $\eta^{(2a)} $ is the vacuum bubble
\begin{align}
	\eta^{(2a)}_{pqrs}
	&=
	\langle 0 \vert \{ p^\dagger q \} \{ r^\dagger s \} \vert  0 \rangle
\end{align}
and $\eta^{(5aL)}_{pqrs}$ are set of all linked diagrams and the 
superscript $L$ is used to represent it. 
Using Wick's theorem, 
\begin{align}
	\{ p^\dagger q \} \{ r^\dagger s \}
	&=
	\{ p^\dagger q r^\dagger s \} 
	+\delta_{qr} \{ p^\dagger  s \}
	-\delta_{ps} \{  r^\dagger  q  \} 
	+\delta_{ps}\delta_{qr}.
\end{align}
Therefore,
\begin{align}
	\eta^{(5aL)}_{pqrs}
	&=
	\eta^{(4b)}_{pqrs}
	+\delta_{qr} \eta^{(4a)}_{ps}
	-\delta_{ps} \eta^{(4a)}_{rq}
	+\delta_{ps}\delta_{qr} \eta^{(3a)}.
\end{align}

\begin{figure*}
  \begin{center}
      % left bottom right top
      %\fbox{\includegraphics[trim=2.5cm 3cm 1.5cm 2.5cm,scale=1.0]{./diagrams_ex_dynamics/table_diag3/table4_1loop.pdf}}  \\
      \includegraphics[trim=2.5cm 3cm 1.5cm 2.5cm,scale=1.0]{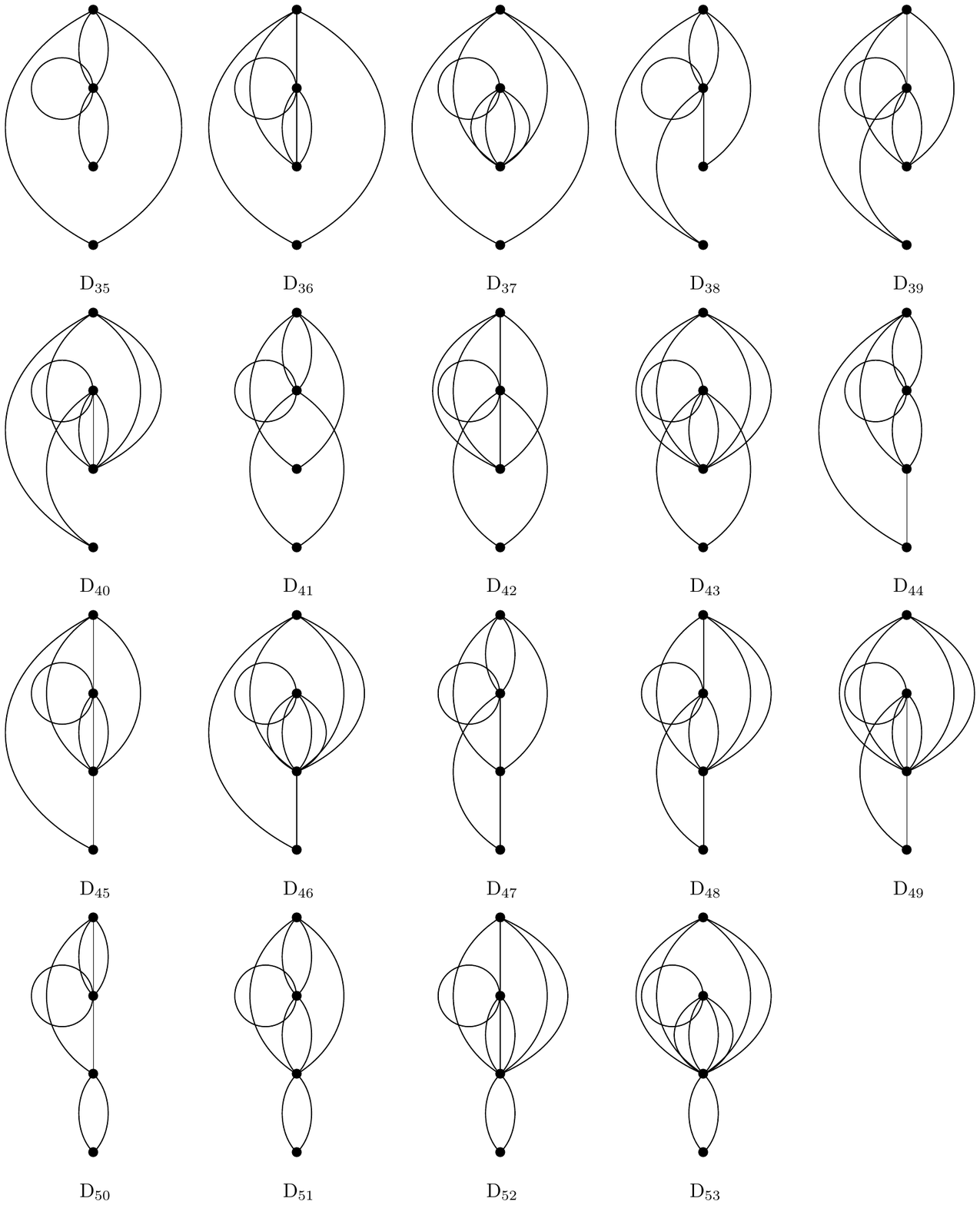}  \\
      \caption{\textrm{1-loop renormalized 4-vertex diagams.}}
    \label{fig:5verta}
  \end{center}
\end{figure*}

\begin{figure*}
  \begin{center}
      % left bottom right top
      %\fbox{\includegraphics[trim=2.5cm 3cm 1.5cm 2.5cm,scale=1.0]{./diagrams_ex_dynamics/table_diag3/table4_2loop.pdf}}  \\
      \includegraphics[trim=2.5cm 3cm 1.5cm 2.5cm,scale=1.0]{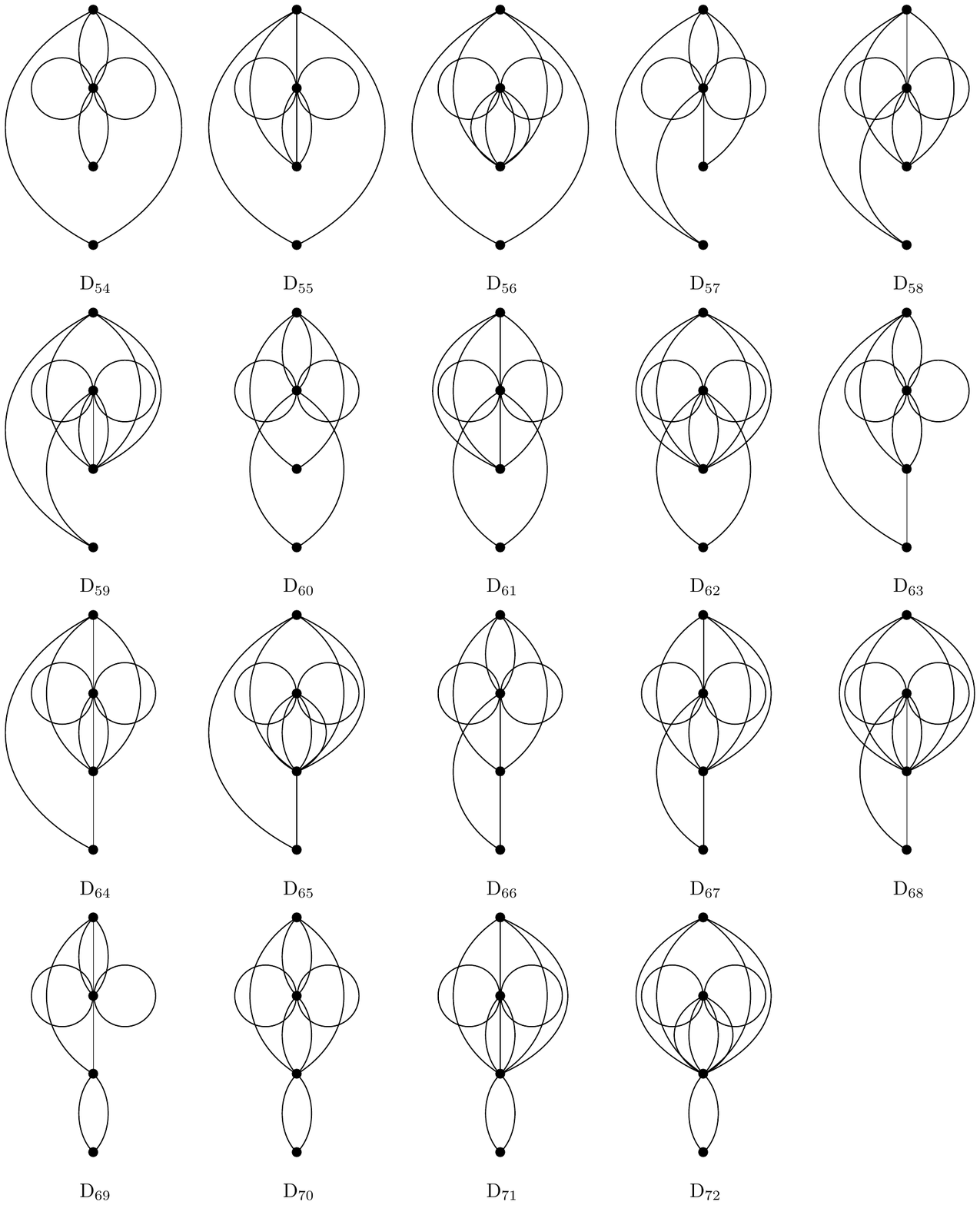}  \\
      \caption{\textrm{2-loop renormalized 4-vertex diagams.}}
    \label{fig:5vertb}
  \end{center}
\end{figure*}

Similarly, the diagrams associated with 
 $\eta^{(5b)}$ can be factored as, 
 \begin{align}
 	\eta^{(5b)}_{pqrsxy}
 	&=
 	\eta^{(1a)}_{xy}  \eta^{(4b)}_{pqrs}
 	+ \eta^{(1b)}_{pqrs}  \eta^{(4a)}_{xy}
 	+ \eta^{(1a)}_{xy} \eta^{(1b)}_{pqrs} \eta^{(3a)}
 	+\eta^{(5bL)}_{pqrsxy} \\
 	\eta^{(5c)}_{pqrstuvw}
 	&=
 	\eta^{(1b)}_{tuvw}  \eta^{(4b)}_{pqrs}
 	+\eta^{(1b)}_{pqrs}  \eta^{(4b)}_{tuvw}
 	+\eta^{(1b)}_{pqrs} \eta^{(1b)}_{tuvw} \eta^{(3a)}
 	+ \eta^{(5cL)}_{pqrstuvw}
 \end{align}
 where, 
 \begin{align}
	\eta^{(1a)}_{pq}
	&=
	\langle 0 \vert \{ p^\dagger q \} \vert  0 \rangle \\
	\eta^{(1a)}_{pqrs}
	&=
	\langle 0 \vert \{ p^\dagger q^\dagger s r \}  \vert  0 \rangle.
\end{align}
In this work, we introduce a renormalization scheme
where all linked 5-vertex diagrams are 
represented as 1-loop and 2-loop renormalized 3-vertex and 4-vertex diagrams. Using this approach, diagrams associated with $\eta^{(5aL)}_{pqrs}$ and $\eta^{(5bL)}_{pqrs}$ are presented in Fig.~\ref{fig:5verta} and Fig.~\ref{fig:5vertb}, respectively.

\section{Evaluation of time-dependent vertex amplitudes}
\label{sec:vertex}
\subsection{Evaluation of time-dependent amplitudes associated with bare 1-body vertex}
In this section, we will evaluate the expression of the 
time-dependent amplitude $A_{pq}(t)$ associated with the
bare 1-body vertex. The equation that defines this amplitude is 
given by the following equation,
\begin{align}
\label{eq:amplitude}
	e^{+iH_0t/\hbar}  V_F(t) e^{-iH_0t/\hbar}
	&=
	\sum_{pq} A_{pq}(t) p^\dagger q. 
\end{align}
We will start by writing the second-quantized (SQ) representation of the 
$V_F(t)$ operator
\begin{align}
	V_F(t) 
	&=
	\sum_{pq} v_{pq}^F(t) p^\dagger q.
\end{align}
Since $v_{pq}^F(t)$ is just a number, we are interested in 
evaluating the SQ operator $e^{+iH_0t/\hbar}  p^\dagger q e^{-iH_0t/\hbar}$.
We will start by inserting identity in this expression,  
\begin{align}
	e^{+iH_0t/\hbar}  p^\dagger q e^{-iH_0t/\hbar}
	&=
	e^{+iH_0t/\hbar}  p^\dagger e^{-iH_0t/\hbar} e^{+iH_0t/\hbar}  q e^{-iH_0t/\hbar}.
\end{align}
The time-dependent creation and annihilation operators are defined as,
\begin{align}
	p^\dagger(t) 
	&=
	e^{+iH_0t/\hbar}  p^\dagger e^{-iH_0t/\hbar}  \\
	q(t) 
	&= 
	e^{+iH_0t/\hbar}  q e^{-iH_0t/\hbar}.
\end{align}
Using BCH expansion,
\begin{align}
 q(t)
 &= 
  q + \frac{it}{\hbar}[q,H_0]
 +\frac{1}{2!} \left( \frac{it}{\hbar}\right)^2 [[q,H_0],H_0] + \dots
\end{align}
Using the results from Eq. \eqref{eq:1_body_comm},
derived in Appendix~\ref{sec:CommutatorIdentities},
\begin{align}
  [p,q^\dagger r] &= \delta_{pq} r
\end{align} 
Therfore,
\begin{align}
	[q,H_0] 
	&= 
	\sum_{p_1q_1} h_{p_1q_1}
	[q,p^\dagger_1,q_1] \\
	&=
	\sum_{p_1q_1} h_{p_1q_1}
	\delta_{qp_1} q_1 \\
	&=
	\sum_{q_1} h_{qq_1} q_{1}.	
\end{align}
Hence, we have the general result,
\begin{align}
	[q,H_0] = \sum_{q_1} h_{qq_1} q_{1}.
\end{align}
Similarly,
\begin{align}
	[[q,H_0],H_0]
	&=
	\sum_{q_1} h_{qq_1} [q_1,H_0]  \\
	&=
	\sum_{q_1 q_2} h_{qq_1} h_{q_1q_2} q_2
\end{align}
We note that the above expression can be 
written in terms of the matrix product
\begin{align}
	\sum_{q_1} h_{qq_1} h_{q_1q_2}
	=
	[\mathbf{h} \mathbf{h}]_{q q_2}
	= [\mathbf{h}^2]_{q q_2}
\end{align}
Therefore, for m-terms expansion,
\begin{align}
	[[q,H_0],\dots, \textrm{m-terms},H_0]
	&=
	\sum_{q_1 q_2 \dots q_m}
	h_{q q_1} h_{q_1 q_2} h_{q_2 q_3} 
	\dots h_{q_{m-1},q_{m}} q_m  \\
	&=
	\sum_{q_m} 
	[\mathbf{h}^m]_{q q_m} q_{m}.
\end{align}
Since $q_m$ is just a summation index,
we can rewrite the expression as,
\begin{align}
\label{eq:comm_series}
[[q,H_0],\dots, \textrm{m-terms},H_0]
	&=
	\sum_{q_1} 
	[\mathbf{h}^m]_{q q_1} q_{1}.
\end{align}
Substituting the above expression in the BCH expansion,
\begin{align}
 q(t)
 &= 
 q + \frac{it}{\hbar} \sum_{q_1} h_{qq_1} q_1
 +\frac{1}{2!} \left( \frac{it}{\hbar}\right)^2 \sum_{q_1} [\mathbf{h}^2]_{qq_1} q_1 +
 \frac{1}{k!} \left( \frac{it}{\hbar}\right)^k \sum_{q_1} [\mathbf{h}^k]_{qq_1} q_1 
  \dots
\end{align}
Combining all the h-terms
\begin{align}
 q(t)
 &= 
 q +
 \sum_{q_1} 
 \left[
  \frac{it}{\hbar}  h_{qq_1} q_1
 +\frac{1}{2!} \left( \frac{it}{\hbar}\right)^2  [\mathbf{h}^2]_{qq_1}  +
 \frac{1}{k!} \left( \frac{it}{\hbar}\right)^k  [\mathbf{h}^k]_{qq_1} 
  \dots
  \right] q_1
\end{align}
Expressing the first term $q$ in terms of $q_1$ using 
Kronecker delta,
\begin{align}
	q = \sum_{q_1} \delta_{qq_1} q_1
\end{align}
we get,
\begin{align}
 q(t)
 &= 
 \sum_{q_1} 
 \left[ \delta_{qq_1} 
  +\frac{it}{\hbar}  h_{qq_1} q_1
 +\frac{1}{2!} \left( \frac{it}{\hbar}\right)^2  [\mathbf{h}^2]_{qq_1}  +
 \frac{1}{k!} \left( \frac{it}{\hbar}\right)^k  [\mathbf{h}^k]_{qq_1} 
  \dots
  \right] q_{1}.
\end{align}
We recognize that the $\delta$ in the above expression
is the element of the identity matrix $\mathbf{I}$.
\begin{align}
 q(t)
 &= 
 \sum_{q_1} 
 \left[ I_{qq_1} 
  +\frac{it}{\hbar}  h_{qq_1} q_1
 +\frac{1}{2!} \left( \frac{it}{\hbar}\right)^2  [\mathbf{h}^2]_{qq_1}  +
 \frac{1}{k!} \left( \frac{it}{\hbar}\right)^k  [\mathbf{h}^k]_{qq_1} 
  \dots
  \right] q_1
\end{align}
We define matrix $\tilde{\mathbf{h}}(t)$ as,
\begin{align}
	\tilde{\mathbf{h}}_A(t)
	&=
	\frac{it}{\hbar} \mathbf{h}.
\end{align}
The subscript $A$ is to remind us that it is an anti-hermitian matrix
\begin{align}
\tilde{\mathbf{h}}_A^\dagger(t)
	&=
	 -\tilde{\mathbf{h}}_A(t).
\end{align}
Using the above definition, the sum in the square brackets can be written in terms of
matrix exponentiation,
\begin{align}
	\sum_{k=0}^{\infty} 
	\frac{1}{k!} \tilde{\mathbf{h}}_A^k(t)
	&= e^{\tilde{\mathbf{h}}_A(t) }
\end{align}
where,
\begin{align}
	\tilde{\mathbf{h}}_A^0 = \mathbf{I}
\end{align}
and $\mathbf{I}$ is identity matrix (and not scalar 1). 
Therefore, the time-development of $q$ is given by,
\begin{align}
 q(t)
  &=
  \sum_{q_1} 
  [e^{\tilde{\mathbf{h}}_A(t) }]_{qq_1} q_1.
\end{align}
Similarly, the time-development of $p^\dagger$ is given by,
\begin{align}
	p^\dagger(t)
	&=
  \sum_{p_1} 
  [e^{-\tilde{\mathbf{h}}_A(t) }]_{pp_1} p_1^{\dagger}.
\end{align}
Therefore,
\begin{align}
	e^{+iH_0t/\hbar}  V_F(t) e^{-iH_0t/\hbar}
	&=
	\sum_{pq} v^F_{pq}(t)  p^\dagger(t) q(t) \\
	&=
	\sum_{pqp_1q_1} 
	v^F_{pq}(t)
	[e^{-\tilde{\mathbf{h}}_A(t) }]_{pp_1} 
	[e^{\tilde{\mathbf{h}}_A(t) }]_{qq_1}
	p_1^\dagger q_{1}.
\end{align}
Using
\begin{align}
	[e^{-\tilde{\mathbf{h}}_A(t) }]^\dagger
	&=
	e^{+\tilde{\mathbf{h}}_A(t) }
\end{align}
\begin{align}
	e^{+iH_0t/\hbar}  V_F(t) e^{-iH_0t/\hbar}
	&=
	\sum_{pqp_1q_1} 
	[e^{+\tilde{\mathbf{h}}_A(t) }]_{p_1p}
	v^F_{pq}(t) 
	[e^{-\tilde{\mathbf{h}}_A(t) }]_{q_1q}
	p_1^\dagger q_1
\end{align}
which is equal to,
\begin{align}
	e^{+iH_0t/\hbar}  V_F(t) e^{-iH_0t/\hbar}
	&=
	\sum_{p_1q_1} 
	[e^{+\tilde{\mathbf{h}}_A(t) }
	\mathbf{v}^F(t) 
	e^{-\tilde{\mathbf{h}}_A(t) }]_{p_1 q_1}
	p_1^\dagger q_{1}.
\end{align}
Comparing to Eq.~\ref{eq:amplitude}, we get the expression for the 
$A$ amplitudes
\begin{align}
	\mathbf{A}(t) = e^{+(it/\hbar)\mathbf{h}} 
	\mathbf{v}^F(t) 
	 e^{-(it/\hbar)\mathbf{h}}.
\end{align}

\subsection{Evaluation of time-dependent amplitudes associated with bare 2-body vertex}
In this section, we will evaluate the expression of the 
time-dependent amplitude $B_{pq}(t)$ associated with the
bare 2-body vertex. The equation that defines this amplitude is 
given by the following equation,
\begin{align}
	e^{+iH_0t/\hbar}  W e^{-iH_0t/\hbar} 
	&=
	\sum_{pqrs} B_{pqrs}(t) p^\dagger q^\dagger s r 
\end{align}
where the 2-body operator is defined as,
\begin{align}
	W = \sum_{pqrs} W_{pqrs} p^\dagger q^\dagger s r.
\end{align}
Using the insertion of identity method used in the 
previous section, we express the above equation 
in terms of time-dependent SQ operators
\begin{align}
	e^{+iH_0t/\hbar}  W e^{-iH_0t/\hbar}
	&=
	\sum_{pqrs} W_{pqrs}
	 p^\dagger(t) q^\dagger(t) s(t) r(t).
\end{align}
Substituting the previously derived expression for 
time-dependent SQ
\begin{align}
	p^\dagger(t)
	&=
  \sum_{p_1} 
  [e^{-(it/\hbar)\mathbf{h}}]_{pp_1} p_1^\dagger 
  =
  \sum_{p_1} 
  [e^{+(it/\hbar)\mathbf{h}}]_{p_1p} p_1^\dagger \\
  s(t)
	&=
  \sum_{s_1} 
  [e^{+(it/\hbar)\mathbf{h}}]_{ss_1} s_1 
  =
    \sum_{s_1} 
  [e^{-(it/\hbar)\mathbf{h}}]_{s_1s} s_1 
\end{align}
we get,
\begin{align}
	e^{+iH_0t/\hbar}  W e^{-iH_0t/\hbar}
	&=
	\sum_{pqrs} W_{pqrs}
	 p^\dagger(t) q^\dagger(t) s(t) r(t) \\
	 &=
	 \sum_{p_1 q_1 r_1 s_1 pqrs} 
	[e^{+(it/\hbar)\mathbf{h}}]_{p_1p}
	[e^{+(it/\hbar)\mathbf{h}}]_{q_1q}	  \\\notag
	&\times W_{pqrs}
	[e^{-(it/\hbar)\mathbf{h}}]_{r_1r}
	[e^{-(it/\hbar)\mathbf{h}}]_{s_1s} \\\notag
	&\times
	p^\dagger_1 q^\dagger_1 s_1 r_{1}. 
\end{align}
The above relationship implies the following 
expression for the $B$,
\begin{align}
 B_{p_1 q_1 r_1 s_1}
 &=
 \sum_{pqrs} 
	[e^{+(it/\hbar)\mathbf{h}}]_{p_1p}
	[e^{+(it/\hbar)\mathbf{h}}]_{q_1q}	  
	W_{pqrs}
	[e^{-(it/\hbar)\mathbf{h}}]_{r_1r}
	[e^{-(it/\hbar)\mathbf{h}}]_{s_1s}.
\end{align}

\section{Results and conclusion}
\label{sec:results}
The main result from this work is the
explicit expressions for the 
time-dependent transition amplitudes 
for generation of 
2e-2h pair from 1e-1h pair
for excited states propagating
in time under the influence of 
external electromagnetic field. 
Up to second-order the 
time-dependent transition amplitude
is given by the following expression,
\begin{align}
	I_F(t_f)
	&=
	I_F^{(0)} t_f 
	+ \int_0^{t_f} dt_1 \, I_F^{(1)}(t_1)  
	+ \int_0^{t_f} d t_1  \int_0^{t_1} d t_2 I_F^{(2)}(t_1,t_2).
\end{align}
Because of the complexity of the equation, a brute-force
approach for the calculation of this expression is computationally 
prohibitive. In this work, we showed that the expressions
for $I_F^{(n)}$ can be separated into a time-dependent 
component and time-independent components. 
We have derived the expression for the time-dependent
components and we show that these quantities 
can be expressed in terms standard matrix-matrix
tensor-tensor contraction terms. 
The extraction of the time-independent components from the
time-propagation equation presents a 
significant computational advantage because the time-independent
component can be evaluated at the start of the calculation and 
can be reused during the course of the time-dependent calculation. 
This strategy dramatically reduces the computational complexity of
for performing such calculations. 
We have also presented the explicit results from the calculation 
of the time-dependent quantities (denoted by $\eta$) 
in terms of the diagrammatic representation. 
\par
One of the key results from this work is the
general treatment of electron correlation 
in the derived result. 
The inclusion of electron-electron 
correlation for the excited state 
is done by the operator $\Omega$ 
in Eq.~\ref{eq:correlationOperator}. In the derivation presented
here, we have not imposed any specific form 
for the electron-correlation operator. 
As a consequence, the set of diagrams 
presented in Fig.~\ref{fig:4verta} and ~\ref{fig:4vertb}, 
is the complete set of diagrams 
associated any form of $\Omega$. 
If $\Omega$ is chosen to be an N-body operator 
like the full-CI or coupled-cluster wave functions,
all the diagrams presented in Fig.~\ref{fig:5verta} and ~\ref{fig:5vertb} will 
contribute to the transition amplitudes. 
However, if $\Omega$ is chosen to be 
a 2-body operator 
only a subset of those diagrams will contribute. 
\par
The complexity and computational cost of the evaluation 
of the diagrams increase with increasing number 
of vertices.  
Out of the 3-vertex, 4-vertex, and 5-vertex diagrams,
the 5-vertex diagrams are most expensive to calculate. 
In this derivation, we have shown that 
a subset of the 5-vertex diagrams can 
be factored into pre-existing 3- and 4-vertex diagrams. 
We also present a renormalization scheme for 
the 5-vertex diagrams by expressing them 
as 1-loop and 2-loop contracted effective 4-vertex diagrams. 
The renormalization method and the factorization of diagrams 
utilizes reusability of pre-computed results and contributes in reducing the overall cost of the calculations.
We envision that the developed method can be used 
for the investigation of time-dependent carrier multiplicity 
in both semiconductor and organic photoactive systems.

\section{Acknowledgments}
We are grateful to National Science Foundation (CHE-1349892) and Syracuse University  for the financial support. AC will also like to thank Prof. Heather Jaeger for insightful discussions about this work.

%\newpage
\section*{Appendix}
\appendix
\section{Commutator identities}
\label{sec:CommutatorIdentities}
The commutator and anticommutator is defined as,
\begin{align}
 [A,B] &= AB - BA \\
 [A,B]_+ &= AB + BA. 
\end{align}
Note that,
\begin{align}
	[B,A]  &= - [A,B] \\
	[B,A]_+ &= [A,B]_{+}. 
\end{align}
The fermionic second-quantized operators 
satisfy the following anticommutation relationships,
\begin{align}
	[p^\dagger , q^\dagger]_+ &= 0 \\
	[p,q]_+ &= 0 \\
	[p^\dagger,q]_+ &= \delta_{pq}.  
\end{align}
This is a well-known identity commutator identity
\begin{align}
	[A,B_1 B_2]
	&=
	B_1 [A,B_2] + [A,B_1] B_2   \\
	[A_1 A_2,B]
	&=
	A_1 [A_2,B] + [A_1,B] A_{2}.
\end{align}
The corresponding anticommutator identity is,
\begin{align}
	[A,B_1 B_2]_+
	&=
	[A,B_1]B_2 + B_1 [A,B_2]_+ \\
	&=
	[A,B_1]_+B_2 - B_1[A,B_2].
\end{align}
The commutator can be written in terms of the
anticommutator as well,
\begin{align}
	[A,B_1 B_2] 
	&=
	[A,B_1]_+ B_2  - B_1[A,B_2]_{+}.
\end{align}
These relationship can be extended to a series of operators
\begin{align}
	[A,B_1 \dots B_N]
	&=
	\sum_{k=1}^N 
	B_1 \dots B_{k-1} [A,B_k] B_{k+1} \dots B_N
\end{align}
\begin{align}
	[A,B_1 \dots B_N]_+
	&=
	\sum_{k=1}^N  (-1)^{k-1}
	B_1 \dots B_{k-1} [A,B_k] _+ B_{k+1} \dots B_{N}.	
\end{align}

The commutation of a single SQ operator with 
1-body operator generates a single SQ operator,
\begin{align}
\label{eq:1_body_comm}
 [p^\dagger, q^\dagger r]  &= -\delta_{pr} q^\dagger \\
 [p,q^\dagger r] &= \delta_{pq} r.
\end{align}

The commutator with two one-body operators 
generate a sum of two one-body operators,
\begin{align}
	[p^\dagger q, r^\dagger s] 
	&=
	\delta_{qr} p^\dagger s 
	-\delta_{ps} r^\dagger q.
\end{align}

The commutator of a one and two-body operator 
generates a sum of two-body operators
\begin{align}
	[p^\dagger q,r^\dagger s m^\dagger n]
	&=
	\delta_{qr} p^\dagger s m^\dagger n
  -\delta_{ps} r^\dagger q m^\dagger n
  +\delta_{qm} r^\dagger s p^\dagger n
  -\delta_{pn} r^\dagger s m^\dagger q.
\end{align}
The general expression for the above results can
be summarized as follows.
The commutator of two 1-body operators
is another 1-body operator,
\begin{align}
	[p_1^\dagger q_1,p_2^\dagger q_2]
	&= 
	\lambda_{pq \dots p_2q_2} p^\dagger q \\
	\lambda 
	&=
	\delta_{q_1 p_2} \delta_{pp_1} \delta_{qq_2}
	-\delta_{p_1 q_2} \delta_{pp_2} \delta_{qq_1}.	
\end{align}
\section{Commutation with 1-body operator}
\begin{align}
	A &= \sum_{p_1 q_1} A_{p_1q_1} p_1^\dagger q_1 \\
	B &= \sum_{p_2 q_2} B_{p_2q_2} p_2^\dagger q_2 
\end{align}
\begin{align}
	[A,B] 
	&= 
	\left[ \sum_{p_1 q_1} A_{p_1q_1} p_1^\dagger q_1,B \right] \\
	&=
	\sum_{p_1 q_1} A_{p_1q_1}
	[p_1^\dagger q_1,B] \\
	&=
	\sum_{p_1 q_1 p_2 q_2} A_{p_1q_1} B_{p_2q_2}
	[p_1^\dagger q_1,p_2^\dagger q_2] 
\end{align}
Using
\begin{align}
	[p_1^\dagger q_1,p_2^\dagger q_2]
	&=
	\delta_{q_1 p_2} p_1^\dagger q_2
	-\delta_{p_1 q_2} p_2^\dagger q_1
\end{align}
We get,
\begin{align}
	[A,B]
	&=
	\sum_{p_1 q_1 p_2 q_2} A_{p_1q_1} B_{p_2q_2}
	[p_1^\dagger q_1,p_2^\dagger q_2]  \\
	&=
	\sum_{p_1 q_1 p_2 q_2} A_{p_1q_1} B_{p_2q_2}
	(	\delta_{q_1 p_2} p_1^\dagger q_2
	-\delta_{p_1 q_2} p_2^\dagger q_1 ) \\
	&=
	\sum_{p_1 q_1 p_2 q_2} A_{p_1q_1} B_{p_2q_2} \delta_{q_1 p_2} p_1^\dagger q_2
	- \sum_{p_1 q_1 p_2 q_2} A_{p_1q_1} B_{p_2q_2} \delta_{p_1 q_2} p_2^\dagger q_1  \\
	&=
	\sum_{p_1 q_2 t } A_{p_1t} B_{tq_2}  p_1^\dagger q_2
	-\sum_{ q_1 p_2 t} A_{tq_1} B_{p_2t}  p_2^\dagger q_1 
\end{align}
Using
\begin{align}
	\sum_t A_{p_1t} B_{tq_2} &= [\mathbf{AB}]_{p_1 q_2} \\
	\sum_t B_{p_2t} A_{tq_1} &= [\mathbf{BA}]_{p_2 q_1}
\end{align}
We get,
\begin{align}
	[A,B]
	&=
	\sum_{p_1 q_2 }  [\mathbf{A}\mathbf{B}]_{p_1 q_2}  p_1^\dagger q_2
	-\sum_{ q_1 p_2 t}  [\mathbf{B}\mathbf{A}]_{p_2 q_1}    p_2^\dagger q_1 
\end{align}
Using general indices, we can write the above expression as,
\begin{align}
	[A,B]
	&=
	\sum_{pq} C_{pq}  p^\dagger q \\
	\mathbf{C} & =  [\mathbf{A},\mathbf{B}].
\end{align}
Therefore, formally we can write that commutator
of two 1-body operators is another 1-body operator,
\begin{align}
	[\hat{A},\hat{B}] = \hat{C}.
\end{align}

\newpage
\bibliography{ex_dynamics}
\end{document}